\documentclass
[aps,preprint,a4paper,tightenlines,nofootinbib,showkeys,showpacs]{revtex4}%
\usepackage{amsfonts}
\usepackage{amsmath}
\usepackage{amssymb}
\usepackage{graphicx}%
\providecommand{\U}[1]{\protect\rule{.1in}{.1in}}
\begin{document}
\title{Adhesive contact of rough surfaces: comparison between numerical calculations
and analytical theories}
\author{G. Carbone$^{1}$, M. Scaraggi$^{1}$, U. Tartaglino$^{2}$}
\affiliation{DIMeG - Politecnico di Bari, v.le Japigia 182, 70126 Bari - Italy}
\affiliation{IFF Forschungszentrum Juelich, 52425 Juelich, Germany}
\begin{abstract}
The authors have employed a numerical procedure to analyze the adhesive
contact between a soft elastic layer and a rough rigid substrate. The solution
of the problem, which belongs to the class of the free boundary problems, is
obtained by calculating the Green's function which links the pressure
distribution to the normal displacements at the interface. The problem is then
formulated in the form of a Fredholm integral equation of the first kind with
a logarithmic kernel, and the boundaries of the contact area are calculated by
requiring that the energy of the system is stationary. The methodology has
been employed to study the adhesive contact between an elastic semi-infinite
solid and a randomly rough rigid profile with a self-affine fractal geometry.
We show that, even in presence of adhesion, the true contact area still
linearly depends on the applied load. The numerical results are then
critically compared with the prediction of an extended version of the
Persson's contact mechanics theory, able to handle anisotropic surfaces, as 1D
interfaces. It is shown that, for any given load, Persson's theory
underestimates the contact area of about 50\% in comparison with our numerical
calculations. We find that this discrepancy is larger than what is found for
2D rough surfaces in case of adhesionless contact. We argue that this
increased difference might be explained, at least partially, by considering
that Persson's theory is a mean field theory in spirit, so it should work
better for 2D rough surfaces rather than for 1D rough surfaces. We also
observe, that the predicted value of separation is in very good agreement with
our numerical results as well as the exponent of the power spectral density of
the contact pressure distribution and of the elastic displacement of the
solid. Therefore, we conclude that Persson's theory captures almost exactly
the main qualitative behavior of the rough contact phenomena.

\end{abstract}

\maketitle

\section{Introduction}

Numerical studies \cite{tartaglino}, \cite{Borri-Brunetto}, \cite{Robbins 1},
\cite{Campana} have shown that, in case of non-adhesive contacts, when an
elastic body is brought into contact with a rough surface the true contact
area increases proportionally to the applied load. To predict such a behavior
two main approaches have been developed: (i) multiasperity contact theories
(originally formulated by Greenwood and Williamson (GW) \cite{Greenwood
Williamson}, \cite{Bush}, \cite{thomas book}, \cite{Greenwood 2006},
\cite{carbone Singolo}) where the contact between the surfaces is modelled as
an ensemble of randomly distributed Hertzian contacts between the asperities,
and (ii) Persson's theory of contact mechanics \cite{Person Rubb Fric JCP},
\cite{PerssonAdhesion} where the probability distribution of the contact
pressure is shown to be governed by a diffusive process as the magnification
at which we observe the interface is increased. The scientific community is
debating about which theory gives the most accurate results. In a previous
paper \cite{CarbBott}\ one of the authors (G.C.) has shown that GW-type
theories predict linearity only for vanishingly small contact areas and load,
whereas as the load is increased the theoretical predictions rapidly deviate
from the asymptotic linearity. This behavior has been shown not to be followed
by Persson's theory, which predicts linearity between contact area and load up
to values of about 15-20\% of nominal contact area, in agreement with some
experimental and numerical results. Numerical calculations by Campa\~{n}\'{a}
et al. \cite{Mueser-Robbins} have shown that Hertzian-type regime, which is
the basis on which GW and similar theories have been developed, occurs only at
relatively small loads, thus indicating the inadequacy of GW-type theories at
higher loads.

As already observed the original version of Persson's theory describes the
interfacial contact pressure through a parabolic partial differential equation
where the diffusivity term is calculated under the approximation that the
Power Spectral Density (PSD)\ of the elastically deformed surfaces is equal to
the PSD of the underlying rough surfaces \cite{Person Rubb Fric JCP},
\cite{persson PSD stress}, i.e. assuming that the diffusive term that one
would obtain in case of full-contact conditions remains exactly the same also
in case of partial contact conditions. The stored elastic energy (or, in case
of sliding of viscoelastic solids, the friction coefficient) is, instead,
calculated assuming that the PSD of the deformed surface is the product of the
PSD\ of the underlying rough surface times the fraction of contact area at the
given resolution. This, in particular, can be shown to be coherently derived
by the theory itself, see Ref. \cite{persson PSD stress}. Of course in full
contact conditions Persson's theory is exact, but in case of partial contact
it has to be possibly verified. For this reason, there is not yet a clear
evidence about the correctness of the factor of proportionality between
contact area and load predicted by Persson's theory. Indeed, there are
numerical investigations \cite{Robbins 1}, \cite{Yang-Persson} of non-adhesive
contacts between rough surfaces, which show that Persson's theory
underestimates the contact area, although its main qualitative prediction
seems in very good agreement with numerical calculations, as proved in Ref.
\cite{Mueser-Robbins}, where Green Function Molecular Dynamics (GFMD)
numerical calculations have been employed to show that the PSD of the deformed
surface has the same power law exponent as predicted by Persson's theory
\cite{persson PSD stress}

In this paper the authors make an attempt to give an additional contribution
in this direction. We extend the analysis to include adhesive interactions at
the interface of the contacting bodies, which become more and more important
as the length scale of observation is decreased and may dominate the contact
behavior of micro- and nano-mechanical and biomechanical systems. We indeed
focus on the adhesive contact between a semi-infinite half space and a
randomly rough surface with roughness in only one direction, and compare our
numerical predictions with the results of Persson's theory. We observe that
the original version of Persson's theory \cite{Person Rubb Fric JCP},
\cite{PerssonAdhesion} was conceived to deal with isotropic surfaces, but in
our case the surfaces is strongly anisotropic, being rough only in one
direction. Therefore, in order to compare theoretical and numerically
calculated data we have employed an extended version of Persson's theory
\cite{carbone Persson anysotropu}, which is able to handle anisotropic
surfaces as in the case of 1D rough surfaces. There are mainly two reasons for
studying a 1D rough surface: (i) first of all one should consider that surface
roughness is characterized by a large number of length scales, which can cover
3-4 decades and even more. Therefore, in order to get physically meaningful
results, one needs to include all the spectral components of the surface
roughness in the analysis. However, increasing the number of length scales
rapidly increases the number of points where the numerical solution has to be
sought, and, in turn, the computation time. This problem is strongly reduced
in case of 1D roughness so that one can include in the analysis more than 3
decades of length scales; (ii) secondly we must also observe that rough
surfaces, encountered in many practical applications, are often strongly
anisotropic mainly as a result of machining and surface treatments (e.g.
unidirectional polished surface which present wear tracks along the polishing
direction, although the resulting roughness is not strictly 1D). Thus, from a
practical point of view, it is also very important to test Persson's
theoretical prediction for anisotropic surfaces.

In the last years scientists have been developing \textit{ad hoc} numerical
methods to treat the problem of contact mechanics between randomly rough
surfaces. Here we would like to recall the methodology proposed by Robbins and
co-workers \cite{Robbins 1}, who developed a Coarse-Graining FEM (CGFEM)
approach, and that conceived by Campa\~{n}\`{a} and M\"{u}ser
\cite{Mueser-Campana}, who have developed a Green's Function Molecular
Dynamics (GFMD) approach to deal with such a problem. Here we employ a
different methodology to deal with adhesive contact. The methodology, already
presented by one of us in Ref. \cite{carbone-thickslab}, is based on a pure
continuum mechanics approach and belongs to the class of Boundary Element
Methods (BEM), since it also makes use of Green's function to solve the
problem. This allows us to reduced the problem to Fredholm integral equation
of the first kind with a logarithmic kernel. We stress that the position of
the edges of each contact patch is not known \textit{a priori }and must be
determined by requiring that the total energy of the system is stationary,
i.e. we are dealing with a free-boundary problem.

The numerical procedure has been designed in such a way to never loose
resolution even when the single contact spot is below the smallest length
scale of observation. We show also that the numerical complexity of the
problem can be strongly reduced since the thermodynamic state of the system
only depends on the size of each contact area and on the pressure distribution
in the contact area. Thus, in our boundary element approach, only the contact
patches need to be discretized (we use an \textit{ad hoc }adaptive grid) and
the solution of the Fredholm equation, which is obtained by means of matrix
inversion, has to be determined only for a limited number of points, i.e. only
those belonging to the true contact area.

\section{The numerical model}

We consider a periodic contact as shown in Figure 1 where an elastic layer of
thickness $d$ is interposed between a flat rigid plate (upper surface) and a
periodically rough rigid substrate with wavelength $\lambda$ (bottom surface).
We assume that the rough surface has roughness in only one direction and is
smooth in the orthogonal direction. Under these conditions the problem at hand
is a periodic plane problem, i.e. the stress, displacement and strain fields
only depends on the $x$ and $y$ coordinates shown in Fig. 2 and are periodic
functions of period $\lambda$. \begin{figure}[ptb]
\begin{center}
\includegraphics[
height=5.77cm,
width=7.99cm
]{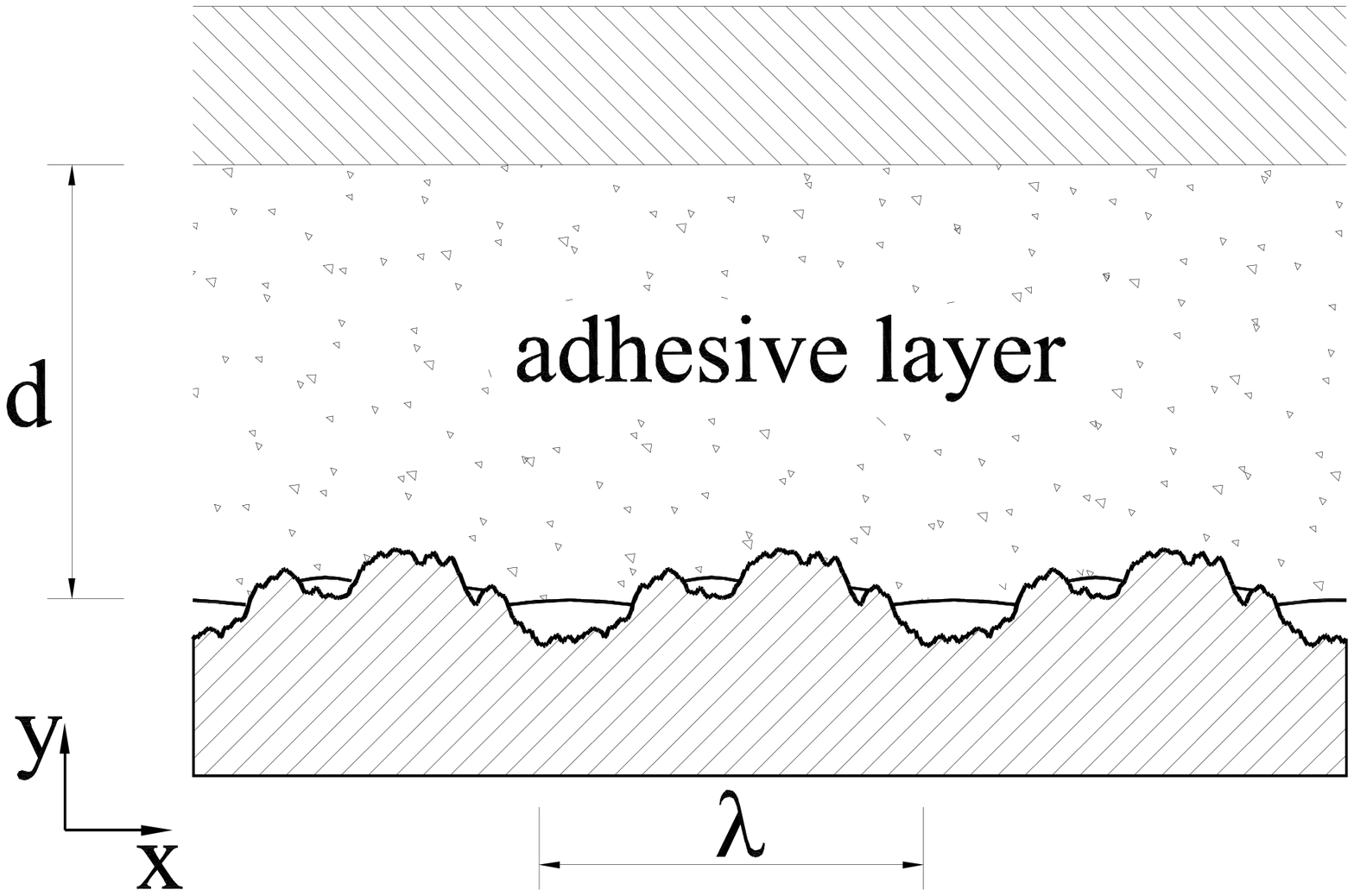}
\end{center}
\caption{An elastic layer of thickness $d$ in adhesive contact with a rough
periodic substrate of wavelength $\lambda$.}%
\label{FIg. Elastic layer}%
\end{figure}Fig. 2 shows, in particular, the total displacement
$u_{\mathrm{tot}}$ of the substrate, the average displacement $u_{\mathrm{m}}$
of boundary of the deformed layer and the penetration $\Delta$ of the rigid
substrate into the elastic slab. These three quantities are shown to satisfy
the following relation%
\begin{equation}
u_{\mathrm{tot}}=\Delta+u_{\mathrm{m}} \label{indenter displacement}%
\end{equation}
\begin{figure}[ptb]
\begin{center}
\includegraphics[
height=6.03cm,
width=10.00cm
]{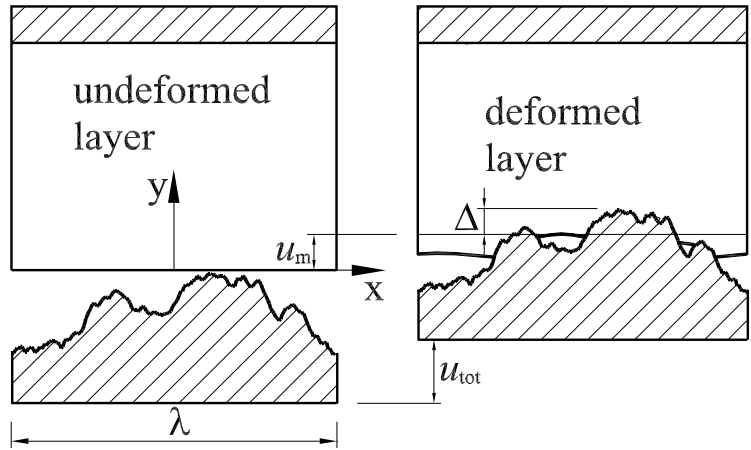}
\end{center}
\caption{The definition of substrate displacement $u_{\mathrm{tot}}$, elastic
layer average displacement $u_{\mathrm{m}}$, and substrate penetration
$\Delta$.}%
\label{Fuig. substrate displacement and penetration}%
\end{figure}We will focus on the pressure distribution $\sigma\left(
x\right)  $ and the displacement $u\left(  x\right)  $ of the elastic solid at
the interface. In Refs. \cite{carbone-thickslab} and \cite{Carbone} G.C. has
shown that the unknown pressure distribution in the contact area
$\mathrm{\Omega}$ can be determined by solving the following Fredholm integral
equation of the first kind with a logarithmic kernel as%
\begin{equation}
-\int_{\mathrm{\Omega}}\mathcal{G}\left(  x-s\right)  \sigma\left(  s\right)
ds=\left[  h\left(  x\right)  -h_{\max}\right]  +\Delta;\qquad x\in
\mathrm{\Omega} \label{integral equation}%
\end{equation}
where $\mathrm{\Omega}=\cup_{i=1}^{L}\left[  a_{i},b_{i}\right]  $ is the
unknown contact domain to be determined as shown below. The quantities $a_{i}$
and $b_{i}$ are the unknown coordinates of $i$-th contact patch with
$a_{i}<b_{i}$ and $i=1,2,...,L$, where $L$ is the unknown number of contacts.
In Eq. (\ref{integral equation}), assuming the elastic layer is infinitely
thick (i.e. $d\rightarrow+\infty$), the kernel is
\begin{equation}
\mathcal{G}\left(  x\right)  =\frac{2\left(  1-\nu^{2}\right)  }{\pi E}%
\log\left[  2\left\vert \sin\left(  \frac{kx}{2}\right)  \right\vert \right]
\label{kernel function}%
\end{equation}
and represents the Green's function of the semi-infinite elastic body under a
periodic loading, i.e. it represents the displacement $u\left(  x\right)
-u_{\mathrm{m}}$ caused by the application of a Dirac comb with peaks
$\delta\left(  x-n\lambda\right)  $ separated by a distance $\lambda$. Here
$E$ and $\nu$ are the Young modulus and the Poisson ratio of the elastic
layer. In Eq. (\ref{integral equation}) and the quantity $h\left(  x\right)  $
represents the heights of the rough profile measured from its mean plane.
Since we are considering a periodic problem $h\left(  x\right)  $ can be
written as Fourier series%
\begin{equation}
h\left(  x\right)  =%
%TCIMACRO{\dsum \limits_{m=1}^{+\infty}}%
%BeginExpansion
{\displaystyle\sum\limits_{m=1}^{+\infty}}
%EndExpansion
h_{m}\cos\left(  mq_{0}x+\phi_{m}\right)  \label{general rough profile}%
\end{equation}
where the fundamental wave vector is $q_{0}=2\pi/\lambda$. Also we have
defined in Eq. (\ref{integral equation}) the quantity $h_{\max}=\max\left[
h\left(  x\right)  \right]  $, which is the maximum height of the substrate
roughness. Once the pressure distribution is known the elastic displacements
at the interface can be easily determined through the equations%

\begin{align}
u\left(  x\right)  -u_{\mathrm{m}}  &  =-\int_{\mathrm{\Omega}}\mathcal{G}%
\left(  x-s\right)  \sigma\left(  s\right)  ds;\qquad x\in\mathrm{D}%
-\mathrm{\Omega}\label{fundsol}\\
u\left(  x\right)  -u_{\mathrm{m}}  &  =h\left(  x\right)  -h_{\max}%
+\Delta;\qquad x\in\mathrm{\Omega}\nonumber
\end{align}
where $\mathrm{D}=\left[  -\lambda/2,\lambda/2\right]  $. Of course for a
infinitely thick layer ($d\rightarrow+\infty$ as in our case) the average
displacement $u_{\mathrm{m}}$ is also infinitely large except when the
$\nu=0.5$, but the difference $u\left(  x\right)  -u_{\mathrm{m}}$ is always
finite \cite{carbone-thickslab}, \cite{Carbone}, and can be interpreted as the
additional elastic displacement of the solid due to the presence of roughness
at the interfaces. In order to close the system of equations we need ad
additional condition to determine the yet unknown contact domain
$\mathrm{\Omega}$. To this end (see also Ref. \cite{carbone-thickslab}), we
first observe that for any penetration $\Delta$, we can calculate the pressure
distribution at the interfaces through Eq. (\ref{integral equation}), and the
interfacial elastic displacement through Eq. (\ref{fundsol}), as functions of
the unknown coordinates $a_{i}$ and $b_{i}$ of the $i$-th contact area. To
calculate the exact values of the quantities $a_{i}$ and $b_{i}$, given
isothermal conditions, we need to find the stationary point of the free
interfacial energy $U_{\mathrm{tot}}\left(  a_{1},b_{1},...,a_{L},b_{L}%
,\Delta\right)  $ of the system for a fixed value of the penetration $\Delta$,
this is the same as requiring that
\begin{equation}
\left(  \frac{\partial U_{\mathrm{tot}}}{\partial a_{i}}\right)
_{\Delta,b_{j}}=0,\qquad\left(  \frac{\partial U_{\mathrm{tot}}}{\partial
b_{i}}\right)  _{\Delta,a_{j}}=0. \label{closure equation}%
\end{equation}
The interfacial energy (see Ref. \cite{carbone-thickslab}) is%
\begin{equation}
U_{\mathrm{tot}}=U_{\mathrm{el}}+U_{\mathrm{ad}} \label{interfacial energy}%
\end{equation}
where we have defined the interfacial elastic energy $U_{\mathrm{el}}$ as the
amount of elastic energy stored in the solid as a consequence of the elastic
deformations caused by the substrate asperities i.e.%
\begin{equation}
U_{\mathrm{el}}\left(  a_{1},b_{1},...,a_{L},b_{L},\Delta\right)  =\frac{1}{2}%
%TCIMACRO{\dsum \limits_{i=1}^{L}}%
%BeginExpansion
{\displaystyle\sum\limits_{i=1}^{L}}
%EndExpansion
\int_{a_{i}}^{b_{i}}\sigma\left(  x\right)  \left[  h\left(  x\right)
-h_{\max}+\Delta\right]  dx. \label{interfacial elastic energy}%
\end{equation}
and the adhesion energy is%
\begin{equation}
U_{\mathrm{ad}}\left(  a_{1},b_{1},...,a_{L},b_{L}\right)  =-\gamma%
%TCIMACRO{\dsum \limits_{i=1}^{L}}%
%BeginExpansion
{\displaystyle\sum\limits_{i=1}^{L}}
%EndExpansion
\int_{a_{i}}^{b_{i}}\sqrt{1+\left[  h^{\prime}\left(  x\right)  \right]  ^{2}%
}dx. \label{adhesion energy general}%
\end{equation}
where $\gamma$ is the Dupr\`{e} energy of adhesion per unit area. Eqs.
(\ref{integral equation}), (\ref{fundsol}) and (\ref{closure equation})
constitute a set of closed equations which allows, for any given penetration
$\Delta$, to determine the coordinates $a_{i}$ and $b_{i}$ of each contact
patch, the pressure distribution at the interface, and all other
thermodynamical quantities. For the numerical implementation the reader is
referred to Ref. \cite{carbone-thickslab}, here we just describe some
numerical techniques which are peculiar to the problem we discuss in this
paper. Let us assume that we know the solution of Eqs.~(\ref{closure equation}%
) for a given penetration $\Delta$: the knowledge of the contact region
$\mathrm{\Omega}=\bigcup_{i=1}^{N}[a_{i},b_{i}]$ is sufficient to fully
characterize the system. Eq.~(\ref{integral equation}) determines the stress
$\sigma(x)$; we solve it iteratively, through the Gauss-Seidel algorithm.
Special care has been taken to guarantee stability and accuracy of the
solution by choosing a suitable sampling grid: the domain $\mathrm{\Omega}$ is
discretized in a non-uniform, adaptive way, so to ensure that more points are
employed close to the edges of the contacts, where the stress distribution,
because of adhesion, presents a square root singularity. As long as the stress
is known, the deformed profile of the elastic layer follows from
Eq.~(\ref{fundsol}). The interfacial energy is then given by
Eqs.~(\ref{interfacial energy}), (\ref{interfacial elastic energy}) and
(\ref{adhesion energy general}).

In summary, at a low level in our numerical implementation of the algorithm
there is the \emph{Solver}, a software code that, given the penetration and
the contact regions, calculates everything else. On top of it we built another
piece of software in charge to adjust the position of the contacts boundaries
${a_{1},b_{1},\ldots,a_{N},b_{N}}$ so to minimize the interfacial energy. We
employed a conjugate gradient method in the version given by Polak and
Ribi\'{e}re \cite{Conj-Grad}. Unfortunately the problem is more complex than a
minimization in a $2N$-dimensional space. Starting form an arbitrary
configuration, some of the contact boundaries can acquire the same value,
meaning that either a contact is detaching or two contacts are coalescing
together into a single bigger one. If this happens, the minimization has to be
restarted in a different number of dimensions.

Furthermore, a constraint must be accounted: in principle we can determine the
configuration corresponding to any penetration and contacts, but we must
impose that in the non-contact regions the elastic layer never intersects the
substrate. For instance, if we consider a physical configuration minimizing
the interfacial energy, and then we increase the penetration pushing the
substrate against the elastic layer, the starting point for the new conjugate
gradient minimization may show an intersection between the elastic layer and
some peaks of the substrate in the non-contact regions. This indicates that a
new contact has to be added before starting the minimization. A specific
procedure inside our software is in charge to detect all the intersections
between elastic layer and substrate, so to enforce the physical constraints of
the problem. Given the penetration $\Delta$, the search of the contact domain
$\mathrm{\Omega}$ that minimizes the interfacial energy is challenging: not
only the $2N$ variables ${a_{1},b_{1},\ldots a_{N},b_{N}}$ are unknown, but
also the number $N$ of contact regions is unknown! The solution of the problem
resorts to conjugate gradient minimization alternated to searches for
intersections between the elastic layer and the block. The minimization
procedure stops when the conjugate gradient ends successfully, i.e.\ it is not
interrupted by a coalescence of detachment of contacts, and the successive
search for intersections confirms that there are no intersections in the
non-contact regions. Although we took special care to guarantee that the
minimization procedure would converge for moderately large variations of
penetration, we observed that the most reliable approach involves many small
increments of penetration starting from 0 (non-contact) up to desired value,
while optimizing of the solution at every intermediate step. The solution of
the contact problem in presence of adhesion is not unique, that is, for the
same penetration more configurations are possible depending on the loading
history. Nonetheless the contact pattern occurring with increasing penetration
is uniquely identified and it is the most suitable solution to represent the
non-adhesive contact in the limit of vanishingly small adhesive bonds. As a
final remark, we observe that this algorithm cannot be used to solve the
problem without adhesion: in this case the solution is still a stationary
point satisfying Eq.~(\ref{closure equation}), but the profile of the elastic
layer in the non-contact regions is always tangent to the substrate near the
crack tips $a_{i}$ and $b_{i}$. An infinitesimal motion of any of the crack
points decreasing the contact region would cause an intersection between
elastic layer and substrate. In other terms, the solution lies always on the
boundary of the subset of $\mathbb{R}^{2N}$ identified by the physical
constraint of non-intersection. The solution would not be a minimum without
such a constraint.

\section{Persson's theory for anisotropic surfaces}

The aim of this paper is to compare the numerical results with analytical
predictions of one of the most promising and also strongly debated theory of
contact mechanics, i.e. the recent theory by Persson \cite{Person Rubb Fric
JCP}, \cite{PerssonAdhesion}. However, the surface we are considering is
strongly anisotropic, in fact it is rough in only one direction and smooth in
the orthogonal direction. The original theory proposed by Persson was,
instead, conceived and developed for perfectly isotropic surfaces. Therefore
in order to carry out the analysis we need to extend this theory to the case
of anisotropic surfaces. This extension has been obtained in Ref.
\cite{carbone Persson anysotropu}. Here we briefly summarize the main
equations. Persson's theory removes the assumption, which is implicit in the
multiasperity contact theories, that the area of real contact is small
compared to the nominal contact area. On the contrary, Persson focuses on the
probability distribution $P\left(  \sigma,\zeta\right)  $ of normal stresses
at the interface, which depends on the magnification at which the contact
interface is observed. To calculate the governing equation of $P\left(
\sigma,\zeta\right)  $, Persson moves from the limiting case of full contact
conditions between a rigid rough surface and an initially flat elastic
half-space \cite{Person Rubb Fric JCP}. In such conditions the PSD\ of the
deformed elastic surface is equal to $C_{\mathrm{2D}}\left(  \mathbf{q}%
\right)  $ where $C_{\mathrm{2D}}\left(  \mathbf{q}\right)  =\left(
2\pi\right)  ^{-2}\int d^{2}x\left\langle h\left(  \mathbf{0}\right)  h\left(
\mathbf{x}\right)  \right\rangle e^{-i\mathbf{q}\cdot\mathbf{x}}$ is the PSD
of the rigid rough substrate (the quantity $h\left(  \mathbf{x}\right)  $ is
the rough substrate height distribution, $\mathbf{x=}\left(  x,y\right)  $ is
the in-plane position vector and the symbol $\left\langle {}\right\rangle $
stands for the ensemble average). Considering that for a perfectly elastic
material the elastic modulus is frequency independent, it can be easily shown
(see Ref. \cite{carbone Persson anysotropu}) that even for the general case of
anisotropic surfaces Persson's theory states that the stress probability
distribution $P\left(  \sigma,\zeta\right)  $ must satisfy the following
relation.%
\begin{equation}
\frac{\partial P\left(  \sigma,\zeta\right)  }{\partial\zeta}=f\left(
\zeta\right)  \frac{\partial^{2}P\left(  \sigma,\zeta\right)  }{\partial
\sigma^{2}} \label{stress diff eq}%
\end{equation}
where the magnification $\zeta=q/q_{0}$, and $\sigma$ is the interfacial
stress in the apparent contact area at the magnification $\zeta$. The
diffusivity function $f\left(  \zeta\right)  =q_{L}G^{\prime}\left(  q\right)
\sigma_{0}^{2}$, where $\sigma_{0}$ is the the average normal stress in the
contact area and $G\left(  q\right)  $ is calculated in full contact
conditions as%
\begin{equation}
G(q)={\frac{1}{8}}\left(  {\frac{E}{1-\nu^{2}}}\right)  ^{2}\frac{1}%
{\sigma_{0}^{2}}\langle\lbrack\nabla h(\mathbf{x})]^{2}\rangle_{q} \label{G}%
\end{equation}
where%
\begin{equation}
\langle\lbrack\nabla h(\mathbf{x})]^{2}\rangle_{q}=\int_{\left\vert
\mathbf{q}^{\prime}\right\vert <q}d^{2}q^{\prime}\ q^{\prime2}C(\mathbf{q}%
^{\prime}) \label{gradiente}%
\end{equation}
is the mean square value of the slope when the surface is observed at the
magnification $\zeta=q/q_{0}$, that is when all harmonic components of the
spectrum with wavevector above $q$ are filtered out. Then, Persson assumes
that Eqs. (\ref{stress diff eq}) and (\ref{G}) hold true also in partial
contact (this is of course an approximation since the PSD\ of the deformed
surface in partial contact conditions cannot be the same as that of the rigid
rough substrate) and to account for partial contact the following initial and
boundary conditions are enforced%
\begin{align}
P\left(  \sigma,1\right)   &  =\delta\left(  \sigma-\sigma_{0}\right)
\nonumber\\
P\left(  -\sigma_{\mathrm{a}},\zeta\right)   &  =0\label{bond and init cond}\\
P\left(  \infty,\zeta\right)   &  =0\nonumber
\end{align}
Here $\sigma_{\mathrm{a}}\left(  \zeta\right)  $ is the tensile stress needed
to cause detachment over a strip of length $2\pi/q$, recalling the Griffith
criterion in plain strain \cite{Griffith} we get%
\begin{equation}
\sigma_{\mathrm{a}}\left(  \zeta\right)  =\left[  \frac{2}{\pi^{2}}E^{\ast
}\gamma_{\mathrm{eff}}\left(  \zeta\right)  q\right]  ^{1/2}
\label{apparent traction stress}%
\end{equation}
where $E^{\ast}=E/\left(  1-v^{2}\right)  $. In Eq.
(\ref{apparent traction stress}) the quantity $\gamma_{\mathrm{eff}}\left(
\zeta\right)  $ is the apparent energy of adhesion at the interface defined as
\cite{PerssonAdhesion}, \cite{carbone mang pers}%
\begin{equation}
-\gamma_{\mathrm{eff}}\left(  \zeta\right)  A\left(  \zeta\right)
=U_{\mathrm{ad}}\left(  \zeta\right)  +U_{\mathrm{el}}\left(  \zeta\right)
\label{gamma eff}%
\end{equation}
where $A\left(  \zeta\right)  $ is the apparent contact area at the
magnification $\zeta$. The calculation of the elastic energy $U_{\mathrm{el}%
}\left(  \zeta\right)  $ must take into account that because of partial
contact conditions the elastic energy stored at the interface is less than
what would be stored in case of full-contact conditions. This is necessarily
true because only where contact occurs the elastic solid conforms the
underlying substrate, whereas outside of the contact regions the elastic
surface is much less deformed. Persson accounts for this fact by assuming
that, when calculating the interfacial elastic energy $U_{el}$, the PSD of the
deformed (initially flat) elastic surface is equal to $C_{\mathrm{2D}}\left(
\mathbf{q}\right)  A\left(  q\right)  /A_{0}$ where $A\left(  q\right)
/A_{0}$ is the fraction of apparent contact area at the length scale $2\pi/q$ i.e.%

\begin{equation}
U_{\mathrm{el}}\left(  \zeta\right)  =\frac{1}{4}E^{\ast}\int_{\left\vert
\mathbf{q}\right\vert >\zeta q_{0}}d^{2}qqC_{\mathrm{2D}}\left(
\mathbf{q}\right)  A\left(  q\right)  \label{elastic energy}%
\end{equation}
This result, that at the beginning was only conjectured by Persson
\cite{Person Rubb Fric JCP}, recently has been demonstrated to directly follow
from Persson's theory itself \cite{persson PSD stress}. Analogously the
adhesion energy $U_{\mathrm{ad}}\left(  \zeta\right)  $ is calculated as%

\begin{equation}
U_{\mathrm{ad}}\left(  \zeta\right)  =-\gamma A\left(  \zeta_{\max}\right)
\int_{0}^{\infty}dx\left(  1+\xi^{2}x\right)  ^{1/2}e^{-x}
\label{energy of adhesion}%
\end{equation}
where
\[
\xi^{2}=\int_{\left\vert \mathbf{q}\right\vert >\zeta q_{0}}d^{2}%
qq^{2}C_{\mathrm{2D}}\left(  \mathbf{q}\right)
\]
and $\zeta_{\max}=q_{1}/q_{0}$, $q_{0}=2\pi/\lambda$, and $q_{1}=2\pi
/\lambda_{1}$ where $\lambda_{1}$ is the shortest length scale of the rough
surfaces. Equations (\ref{stress diff eq}, \ref{bond and init cond},
\ref{apparent traction stress}, \ref{gamma eff}, \ref{elastic energy} and
\ref{energy of adhesion}) can be solved to calculate the stress probability
distribution $P\left(  \sigma,\zeta\right)  $ and hence the apparent contact
area $A\left(  \zeta\right)  /A_{0}$ as a function of the magnification
$\zeta$ \cite{PerssonAdhesion}, \cite{carbone mang pers}
\begin{equation}
\frac{A\left(  \zeta\right)  }{A_{0}}=\int_{-\sigma_{a}\left(  \zeta\right)
}^{\infty}P\left(  \sigma,\zeta\right)  d\sigma\label{contact area general}%
\end{equation}
The separation $s=h_{\max}-\Delta$ can be calculated observing that the change
of total interfacial energy $U_{\mathrm{tot}}=U_{\mathrm{el}}+U_{\mathrm{ad}}$
must be equal to the work done by the applied load, i.e.%
\begin{equation}
dU_{\mathrm{tot}}=\sigma_{0}A_{0}d\Delta=-\sigma_{0}A_{0}ds
\label{diff eq. penetration}%
\end{equation}
which gives%
\begin{equation}
s=\int_{\sigma_{0}}^{+\infty}\frac{1}{A_{0}\sigma_{0}^{\prime}}\frac
{dU_{\mathrm{tot}}}{d\sigma_{0}^{\prime}}d\sigma_{0}^{\prime}
\label{equation for the separation}%
\end{equation}
In case of non-adhesive contact i.e. when $\gamma=0$, Persson has shown that
\begin{equation}
P\left(  \sigma,\zeta\right)  =\frac{1}{2\left(  \pi G\right)  ^{1/2}}\left[
e^{-\left(  \sigma-\sigma_{0}\right)  ^{2}/4G}-e^{-\left(  \sigma+\sigma
_{0}\right)  ^{2}/4G}\right]  \label{prob dens distrib stress}%
\end{equation}
and Eq. (\ref{contact area general}) simply becomes%
\begin{equation}
\frac{A\left(  \zeta\right)  }{A_{0}}=\operatorname{erf}\left(  {\frac
{1}{2\surd G(\zeta)}}\right)  \label{apparent contact area}%
\end{equation}
The above formulation holds true also for anisotropic surfaces, in particular
if the substrate has roughness in only one direction, e.g. along the
$x$-direction, we get $\left\langle \left[  \nabla h(\mathbf{x})\right]
^{2}\right\rangle _{q}=\left\langle \left(  \partial h/\partial x\right)
^{2}\right\rangle _{q}$. In such a case we have $h(\mathbf{x})=h(x)$ and one
can easily show that%
\begin{equation}
C_{\mathrm{2D}}(\mathbf{q})=C(q_{x})\delta(q_{y}) \label{2D 1D PSD}%
\end{equation}
where $C\left(  q\right)  =\left(  2\pi\right)  ^{-1}\int dx\left\langle
h\left(  0\right)  h\left(  x\right)  \right\rangle e^{-iqx}$ is the PSD of
the $x$-profile of the surface. Using Eq. (\ref{2D 1D PSD}) one simply obtains%
\begin{equation}
\left\langle \left(  \partial h/\partial x\right)  ^{2}\right\rangle _{q}%
=\int_{-q}^{q}dq_{x}q_{x}^{2}C(q_{x}) \label{1D PSD gradient}%
\end{equation}
Eq. (\ref{contact area general}) gives the apparent contact area at the
resolution $\lambda\left(  q\right)  =2\pi/q$ as a function of the applied
load $\sigma_{0}$. However, we are interested in calculating the real contact
area, which can be obtained by replacing $\zeta$ with $\zeta_{\max}$.

\section{Rough profile generation}

In order to carry out the numerical simulations and compare the results with
the theoretical predictions, we need to numerically generate a rough profile.
We have opted for a fractal self affine rough profile. For any self affine
fractal profile $h\left(  x\right)  $ the statistical properties are invariant
under the transformation%
\begin{equation}
x\rightarrow tx;\qquad h\rightarrow t^{H}h \label{self affine transfomation}%
\end{equation}
in such a case it can be shown that the PSD\ of the profile is%
\begin{equation}
C\left(  q\right)  =C_{0}\left(  \frac{\left\vert q\right\vert }{q_{0}%
}\right)  ^{-\left(  2H+1\right)  } \label{PSD}%
\end{equation}
where $H$ is the Hurst exponent of the randomly rough profile, and is related
to the fractal dimension $D_{f}=2-H$. In order to carry out the numerical
calculations we have utilized a periodic profile with Fourier components up to
the value $q_{1}=Nq_{0}$ (in this case $\zeta_{1}=N$). However we need to
determine the amplitudes $h_{m}$ and the phases $\phi_{m}$ of the harmonic
terms [see Eq. (\ref{general rough profile})]. It can be shown that in order
to satisfy the translational invariance of the profile's statistical
properties (which implies that the autocorrelation function satisfies the
relation $\left\langle h\left(  x^{\prime}\right)  h\left(  x^{\prime
}+x\right)  \right\rangle =\left\langle h\left(  0\right)  h\left(  x\right)
\right\rangle $), it is enough to assume that the random phases $\phi_{m}$ are
uniformly distributed on the interval $\left[  -\pi,\pi\right[  $. In such a
case the autocorrelation of the profile takes the form
\begin{equation}
\left\langle h\left(  x^{\prime}\right)  h\left(  x^{\prime}+x\right)
\right\rangle =%
%TCIMACRO{\dsum \limits_{m=1}^{N}}%
%BeginExpansion
{\displaystyle\sum\limits_{m=1}^{N}}
%EndExpansion
\frac{\left\langle h_{m}^{2}\right\rangle }{2}\cos\left(  mq_{0}x\right)
\label{autocorr function}%
\end{equation}
Now we need to calculate the quantities $\left\langle h_{m}^{2}\right\rangle
$. To this purpose let us calculate the PSD of the periodic profile of Eq.
(\ref{general rough profile}). By using the definition we get
\begin{equation}
C\left(  q\right)  =\left(  2\pi\right)  ^{-1}%
%TCIMACRO{\dsum \limits_{m=1}^{N}}%
%BeginExpansion
{\displaystyle\sum\limits_{m=1}^{N}}
%EndExpansion
\int dx\frac{\left\langle h_{m}^{2}\right\rangle }{2}\cos\left(
mq_{0}x\right)  e^{-iqx}=%
%TCIMACRO{\dsum \limits_{m=1}^{N}}%
%BeginExpansion
{\displaystyle\sum\limits_{m=1}^{N}}
%EndExpansion
\frac{1}{4}\left[  \left\langle h_{m}^{2}\right\rangle \delta\left(
q-mq_{0}\right)  +\left\langle h_{m}^{2}\right\rangle \delta\left(
q+mq_{0}\right)  \right]  \label{PSD numerical}%
\end{equation}
from which it follows that%
\[
C\left(  -mq_{0}\right)  =C\left(  mq_{0}\right)  =C_{m}=\frac{\left\langle
h_{m}^{2}\right\rangle }{4}\delta\left(  0\right)
\]
Using Eq. (\ref{PSD}) and observing that $\ C_{0}=\left\langle h_{1}%
^{2}\right\rangle \delta\left(  0\right)  /4$, one then obtains%
\begin{equation}
\left\langle h_{m}^{2}\right\rangle =\left\langle h_{1}^{2}\right\rangle
m^{-\left(  2H+1\right)  } \label{amplitudes}%
\end{equation}
Hence, the quantity $\left\langle h_{m}^{2}\right\rangle $ can be determined
once known $\left\langle h_{1}^{2}\right\rangle $ and the Hurst exponent of
the surface. \begin{figure}[ptb]
\begin{center}
\includegraphics[
height=6.54cm,
width=10.00cm
]{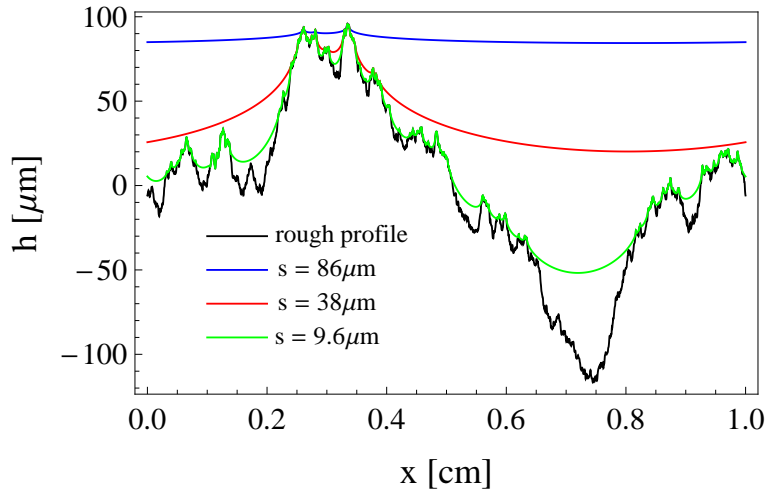}
\end{center}
\caption{The deformed shape of the elastic body at three different
separations, $s=86\mathrm{\mu m}$ (blue), $s=48\mathrm{\mu m}$ (red), and
$s=9.6\mathrm{\mu m}$ (green). The rough rigid substrate profile is shown in
black color.}%
\label{Fig. shape of deformed body}%
\end{figure}\ Now observe from Eq. (\ref{autocorr function}) that
$\left\langle h\left(  x\right)  ^{2}\right\rangle =\sum_{m=1}^{N}\left\langle
h_{m}^{2}\right\rangle /2$, and using Eq. (\ref{amplitudes}) one obtains%
\begin{equation}
\left\langle h\left(  x\right)  ^{2}\right\rangle =\frac{\left\langle
h_{1}^{2}\right\rangle }{2}\sum_{m=1}^{N}m^{-\left(  2H+1\right)  }
\label{rms}%
\end{equation}
Therefore, if one knows the \textit{rms} roughness of the surfaces
$h_{\mathrm{rms}}=\sqrt{\left\langle h\left(  x\right)  ^{2}\right\rangle }$
one can calculate $\left\langle h_{1}^{2}\right\rangle $ and therefore all the
other quantities $\left\langle h_{m}^{2}\right\rangle $. However to completely
characterize the rough profile we still need the probability distribution of
the amplitudes $h_{m}$. There are several choices, however the simplest
assumption, as suggested by Persson et al. in Ref. \cite{Persson Roughness},
is that the probability density function of $h_{m}$ is just a Dirac's delta
function centered at $\left[  4C_{m}/\delta\left(  0\right)  \right]
^{1/2}\approx2\sqrt{2\pi}C_{m}^{1/2}/L^{1/2}$, i.e.
\begin{equation}
p\left(  h_{m}\right)  =\delta\left(  h_{m}-2\sqrt{\frac{2\pi}{L}}C_{m}%
^{1/2}\right)  \label{PDF delta}%
\end{equation}
where we have used that $\delta\left(  q=0\right)  \approx L/\left(
2\pi\right)  $. In can be shown \cite{Persson Roughness} that this choice
guarantees also that the random profile $h\left(  x\right)  $ has a Gaussian
random distribution.

\section{Results}

We assume that the elastic block is a soft perfectly elastic material with
elastic modulus $E=1\mathrm{MPa}$ and Poisson's ratio $\nu=0.5$, i.e. we
assume that the material is incompressible. We assume also that the change of
surface energy upon contact between the two surfaces (i.e. the Dupr\`{e}
energy of adhesion) is $\gamma=0.03\mathrm{J/m}^{2}$. Calculations have been
carried out for 11 different realizations of a rough self-affine fractal 1D
profile. \begin{figure}[ptb]
\begin{center}
\includegraphics[
height=6.61cm,
width=10.00cm
]{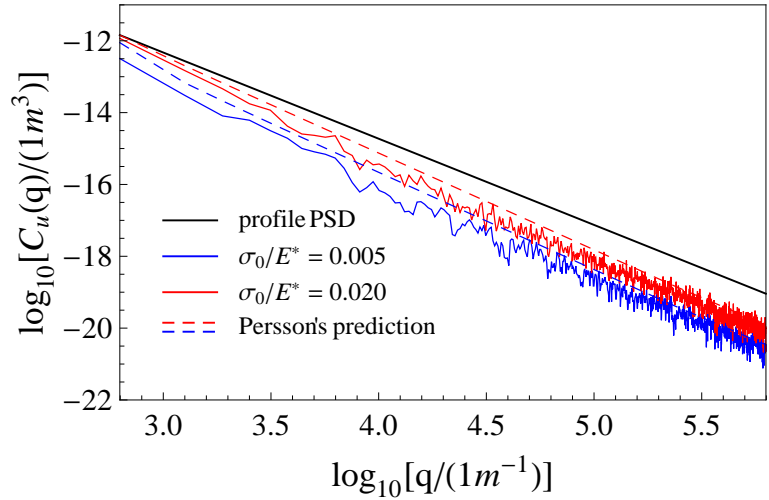}
\end{center}
\caption{The PSD\ $C_{u}\left(  q\right)  $ of the rigid substrate profile
(black solid line) compared to that of the deformed shape of the elastic body
for two different dimensionless loads $\sigma_{0}/E^{\ast}=0.005$ (blue), and
$\sigma_{0}/E^{\ast}=0.020$ (red). Solid lines refer to numerical calculations
whereas dashed lines refer to Persson's theory. The agreement between
Persson's theory and numerical calculated predictions is qualitatively very
good. Indeed the slope of the PSD\ predicted by Persson (in the mid range of
$q$-vectors where the influence of adhesion is negligible) is almost the same
as the numerically calculated one (see text): in both cases the PSD of the
elastically deformed solid is $C_{u}\left(  q\right)  \approx q^{-\left(
2+H\right)  }$.}%
\label{FIg. PSD u}%
\end{figure}\ The profile has a fractal dimension $D_{f}=1.3$ (i.e. the Hurst
coefficient is $H=0.7$), with root mean square roughness $\left\langle
h^{2}\right\rangle ^{1/2}=50\mathrm{\mu m}$. The self-affine profiles have
spectral components in the range $q_{0}<q<q_{1}$. We have used $\lambda
=2\pi/q_{0}=0.01\mathrm{m}$ and $q_{1}=10^{3}q_{0}$. The numerical
calculations have been carried out for different values of the separations
$s=h_{\max}-\Delta$, which is defined as the distance between the mean plane
of the deformed surface and the mean plane of the rough surfaces. In Fig.
\ref{Fig. shape of deformed body} we show three different shapes of the
deformed profiles at three different values of the separation:
$s=86\mathrm{\mu m}$ (blue), $s=38\mathrm{\mu m}$ (red), and $s=9.6\mathrm{\mu
m}$ (green). The black line instead represents the rigid rough substrate
profile. A deeper analysis of the figure shows that, not depending on the
separation $s$, full contact occurs between the elastic block and the short
wave length corrugation of the rough rigid profile. \begin{figure}[ptb]
\begin{center}
\includegraphics[
height=6.67cm,
width=10.00cm
]{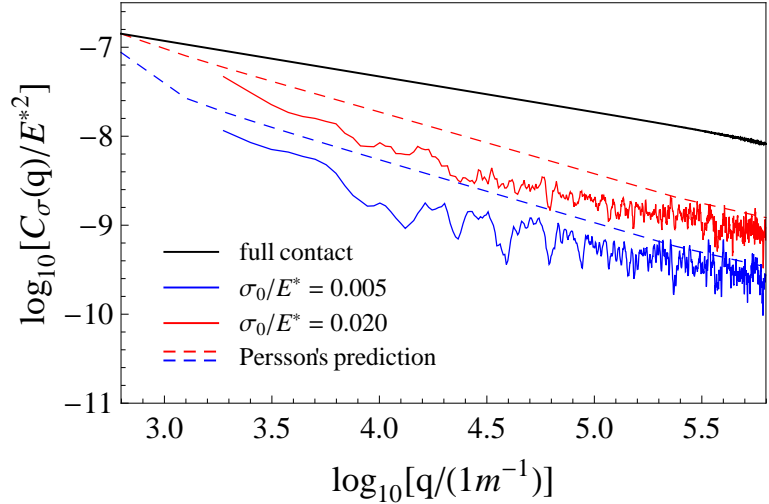}
\end{center}
\caption{The quantity\ $C_{\sigma}\left(  q\right)  /E^{\ast2}$ for full
contact conditions (black solid line) and for two different dimensionless
loads $\sigma_{0}/E^{\ast}=0.005$ (blue), and $\sigma_{0}/E^{\ast}=0.020$
(red). Solid lines refer to numerical calculations whereas dashed lines refer
to Persson's theory. Also in this case the agreement between Persson's theory
and numerical calculated predictions is qualitatively very good, at least in
the mid range of $q$-vectors where the influence of adhesion is negligible:
both calculations predict $C_{\sigma}\left(  q\right)  \approx q^{-H}$.}%
\label{Fig. PSD stress}%
\end{figure}\ This is in agreement with some theoretical arguments
\cite{PerssonAdhesion}, \cite{carbone mang pers} which predict that this
situation should occur when the Hurst exponent of the rough profile is larger
than 0.5, as in our case. This is also confirmed in Fig. \ref{FIg. PSD u}
which shows the PSD of the rough surface and that of the deformed elastic
surface as a function of the wave-vector $q$ (in a log-log diagram) for two
values of the applied stress $\sigma_{0}/E^{\ast}=0.005$ (blue), and
$\sigma_{0}/E^{\ast}=0.020$ (red), where $E^{\ast}=E/\left(  1-\nu^{2}\right)
$. Indeed, we observe that for large $q$-vectors the PSD $C_{u}\left(
q\right)  =\left(  2\pi\right)  ^{-1}\int dx\left\langle u\left(  0\right)
u\left(  x\right)  \right\rangle e^{-iqx}$ of the numerically calculated
deformed profile $u\left(  x\right)  $ becomes almost perfectly parallel to
the PSD $C\left(  q\right)  $ of the rigid substrate: this means that the
spectral content of the deformed body profile at short wavelength is just the
same as that of the rough rigid profile, and therefore that full contact
occurs between the elastic body and the substrate at short wavelengths. Fig.
\ref{FIg. PSD u}, shows also, as expected, that, as the load is increased, the
quantity $C_{u}\left(  q\right)  $ continuously approaches the PSD\ $C\left(
q\right)  $ of the rigid rough profile and must become equal to $C\left(
q\right)  $ at very high loads, i.e. when full contact occurs.
\begin{figure}[ptb]
\begin{center}
\includegraphics[
height=6.22cm,
width=10.00cm
]{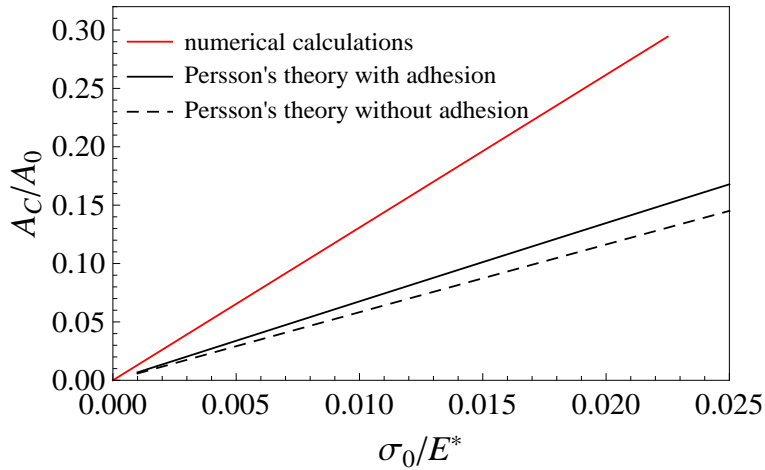}
\end{center}
\caption{The true contact area $A\left(  \zeta_{\max}\right)  $ in units of
the the nominal contact area $A_{0}$ as a function of the dimensionless
applied load $\sigma_{0}/E^{\ast}$. Numerical predictions are in red, whereas
Persson's theoretical calculations are in black colour. Observe that both
numerical calculations and Persson's theory predict linearity between contact
area and load. However Persson's theory predicts a coefficient of
proportionality which is about half of the numerically calculated one. The
dashed line represents Persson's calculations in absence of adhesion
interactions. Notice that because of the low amount of adhesion the curve does
not differ significantly from the solid line.}%
\label{Fig. Area vs Press}%
\end{figure}\ In Fig. \ref{FIg. PSD u} we also compare the numerically
calculated PSD of the deformed surface (solid line) with Persson's theoretical
predictions (dashed line). We first observe that there is a non negligible
shift between Persson's results and our numerically calculated ones. However
the two curves run almost perfectly parallel especially in the mid range of
wavevectors where the best fit of our numerical results gives $C_{u}\left(
q\right)  \approx q^{-2.73}$. The value $-2.73$ is very close to the value
$-2.7$ predicted by Persson. Indeed, Persson's theory relies on the assumption
that the PSD of the deformed surfaces $C_{u}\left(  q\right)  =C\left(
q\right)  A\left(  q\right)  /A_{0}$, where $C\left(  q\right)  $ is the PSD
of the rough substrate. Now, in case of self affine fractal surfaces, using
Eq. (\ref{PSD}) and observing that if one neglects adhesion (this is correct
in the mid range of wavevectors $q$) Eqs. (\ref{G}) and
(\ref{apparent contact area}) give $A\left(  q\right)  /A_{0}\approx q^{H-1}$,
one obtains from Persson's theory that $C_{u}\left(  q\right)  \approx
q^{-\left(  2+H\right)  }$. Being, in our case, $H=0.7$, we have $C_{u}\left(
q\right)  \approx q^{-2.7}$in perfect agreement with our numerical
calculations. \begin{figure}[ptb]
\begin{center}
\includegraphics[
height=6.91cm,
width=10.00cm
]{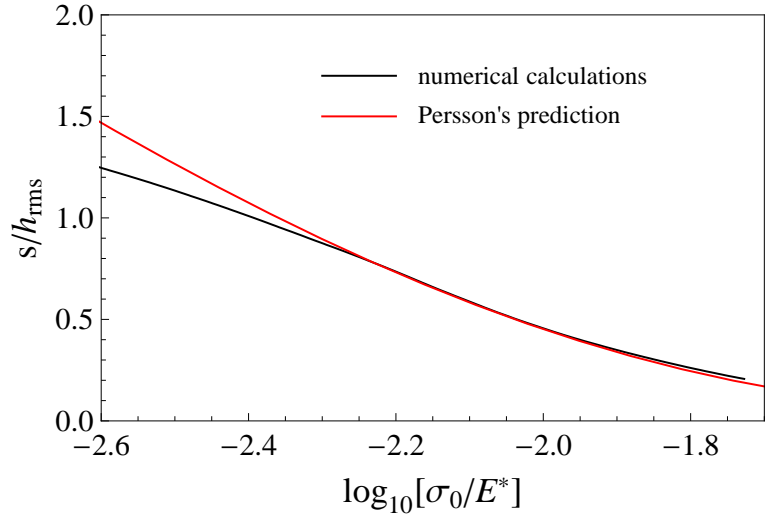}
\end{center}
\caption{The separation $s$ in units of roughness $h_{\mathrm{rms}}$ as a
function of the dimensionless applied load $\sigma_{0}/E^{\ast}$ in a
log-linear scale. The agreement between numerical data (black curve) and the
Persson's theoretical predictions (red curve) is almost perfect. The numerical
calculations deviate from theoretical predictions only at small loads, since
our system is a finite system so a finite value of the separation $s$
necessarily exists at which contact area goes to zero and therefore also the
applied load. Persson's theory instead has been developed for infinite
systems. In this case indeed the rough surface has arbitrarily many and
arbitrarily high asperities, which always allows the contact between the two
solids to occur for arbitrarily large surface separations.}%
\label{FIg. Penetrazione}%
\end{figure}\ Same conclusions can be found if one observes Fig.
\ref{Fig. PSD stress} which shows in a log-log diagram the power spectral
density $C_{\sigma}\left(  q\right)  =\left(  2\pi\right)  ^{-1}\int
dx\left\langle \sigma\left(  0\right)  \sigma\left(  x\right)  \right\rangle
e^{-iqx}$ of the stress distribution at the interface in units of $E^{\ast2}$.
Of course this is not unexpected since the $C_{\sigma}\left(  q\right)  $ and
$C_{u}\left(  q\right)  $ are related each-other through $C_{\sigma}\left(
q\right)  =\frac{1}{4}E^{\ast2}q^{2}C_{u}\left(  q\right)  $ \cite{persson PSD
stress}, \cite{carbone Persson anysotropu}. Using that $C_{u}\left(  q\right)
\approx q^{-\left(  2+H\right)  }$, and assuming the adhesion interaction is
not important (which occurs in the mid range of wavevectors $q$), one obtains,
$C_{\sigma}\left(  q\right)  \approx q^{-H}$. This is indeed confirmed in Fig.
\ref{Fig. PSD stress} which shows that Persson's prediction (dashed lines) and
numerical calculations (solid lines) run parallel to each other in low-mid
range of $q$-vectors. However, as $q$ is increased the numerical calculated
PSD $C_{\sigma}\left(  q\right)  $ rapidly changes its slope.
\begin{figure}[ptb]
\begin{center}
\includegraphics[
height=6.51cm,
width=10.00cm
]{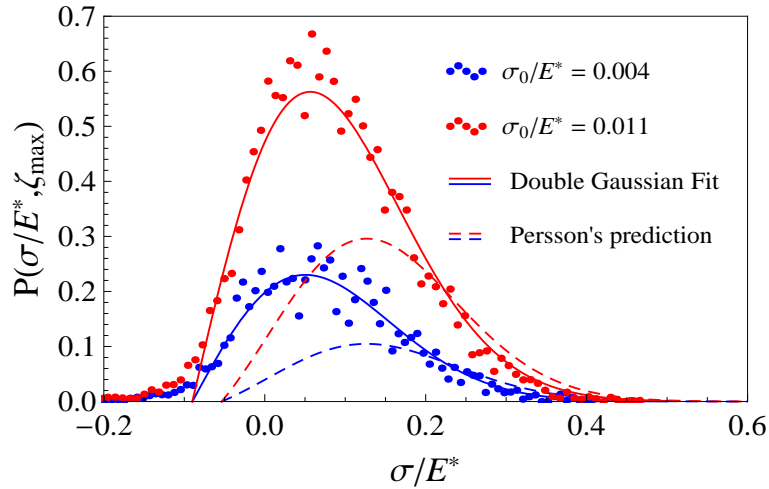}
\end{center}
\caption{The probability function $P\left(  \sigma,\zeta_{\max}\right)  $ of
interfacial pressure distribution $\sigma\left(  x\right)  $. Points are
numerical predictions whereas dashed lines are Persson's results. The trend is
qualitatively the same, although it is quantitatively different. The reason
for such a difference is that numerical calculations and Persson's theory do
not predict the same value of the contact area for any given applied load (see
Fig. \ref{Fig. Area vs Press}). Notice that, the tail of numerically
calculated probability distribution at negative loads is an effect of the
adhesion interaction which has been introduced only through the surface
energy, i.e. with the inclusion of an interaction force with an
infinitesimally short range. We also present a best fit based on double
Gaussian probability distribution (see text).}%
\label{FIg_PDF}%
\end{figure}\ This, in turn, becomes almost equal to that of the PSD of
interfacial normal stress distribution that would be obtained in full contact
conditions (solid black line in Fig. \ref{Fig. PSD stress}), and confirms,
what we have already observed above, i.e. that because of adhesion the short
wavelength roughness of the underlying rigid profile is in full contact with
the elastic body. Figure \ref{Fig. Area vs Press} shows the true contact area
$A\left(  \zeta_{\max}\right)  /A_{0}$ vs. the dimensionless load $\sigma
_{0}/E^{\ast}$ calculated through Persson's theory for adhesive contact and
the one computed by our numerical code. Figure \ref{Fig. Area vs Press}
confirms the linearity between contact area and load, however it also shows a
significant disagreement between Persson's theory and our numerical
calculations. In particular, Persson's theory predicts a contact area which is
about 50\% less than that calculated with our numerical code, in agreement
with some molecular dynamics simulations \cite{Yang-Persson}, where Persson
himself has found a difference between the numerically calculated contact area
and the theoretical value of about 30\% for a two-dimensional rough surface.
The dashed curve in Fig. \ref{Fig. Area vs Press} represents Persson's
theoretical predictions when adhesion is not included in calculations. As
expected adhesive interactions lead to an increase of the contact area.
\begin{figure}[ptb]
\begin{center}
\includegraphics[
height=6.59cm,
width=10.00cm
]{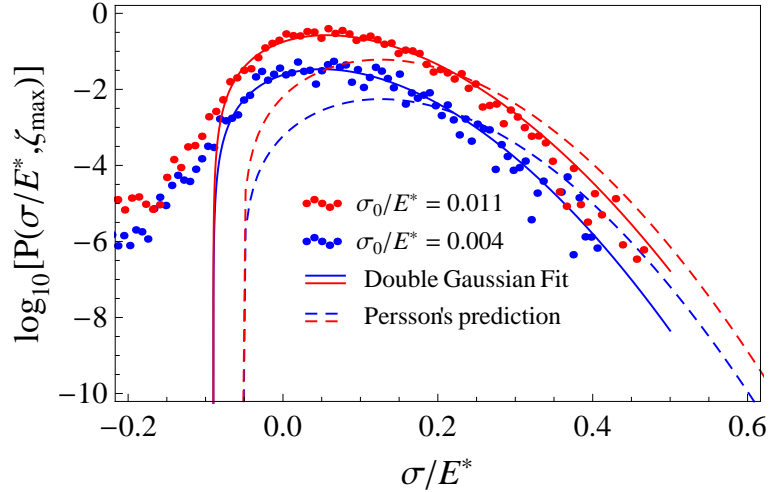}
\end{center}
\caption{The logarithm of the probability function $P\left(  \sigma
,\zeta_{\max}\right)  $. Points are numerical predictions whereas dashed lines
are Persson's results. We observe that the tail of the probability
distribution at large values of $\sigma$ follow exactly a Gaussian
distribution, whereas the tail obtained for negative value of $\sigma$ is not
Gaussian.}%
\label{Fig. PDF-LOG}%
\end{figure}However, being in our case the amount of adhesion energy
relatively small, this effect is only marginal. We observe that the large
discrepancy between Persson's predictions and our numerical calculations may
be explained, at least partially, by the fact that Persson's theory is a mean
field theory in spirit, so it should work better for 2D rough surfaces rather
than for 1D rough surfaces. Fig. \ref{FIg. Penetrazione} shows the
dimensionless separation $s/h_{\mathrm{rms}}$ as a function of the
dimensionless load $\sigma_{0}/E^{\ast}$. Numerical predictions (black line)
are compared to Persson's ones. The agreement is very good except at lower
applied loads. Indeed at small loads the numerically calculated separation
drops off faster than predicted by Persson's theory. The same effect has also
been observed in molecular dynamics calculations \cite{Yang-Persson}. The
explanation for this behavior is that numerical calculations have been carried
out for a finite system, whereas Persson's theory is for an infinite system.
Indeed an infinite system has many (arbitrarily) high asperities, which always
allows the contact between the two solids to occur for arbitrarily large
surface separations. But a finite system has asperities with height below some
finite length $h_{\max}$, and for $u>h_{\max}$ no contact occurs between the
solids and $\sigma_{0}\rightarrow0$. Fig. \ref{FIg_PDF} shows the probability
density function $P\left(  \sigma,\zeta_{\max}\right)  $ of interfacial normal
stress distribution $\sigma\left(  x\right)  $ at the highest magnification.
We observe that Persson's predictions and numerical data agree only at a
qualitative level, but strongly deviates from a quantitative point of view.
The reason of this quantitative disagreement can be easily understood if one
recalls Eq. (\ref{contact area general}) which states that the integral of the
$P\left(  \sigma,\zeta_{\max}\right)  $ must be proportional to the true
contact area $A\left(  \zeta_{\max}\right)  $ and considers that Persson's
theory predicts a contact area smaller by a factor $\approx1/2$, if compared
to numerical calculations. We also present with solid lines the best fit
obtained with the double Gaussian distribution
\begin{equation}
P\left(  \sigma,\zeta_{\max}\right)  =\frac{1}{2\left(  \pi G\right)  ^{1/2}%
}\left[  e^{-\left(  \sigma-\sigma_{0}+\sigma_{\mathrm{a}}\right)  ^{2}%
/4G}-e^{-\left(  \sigma+\sigma_{0}+\sigma_{\mathrm{a}}\right)  ^{2}%
/4G}\right]  \label{final double gaussian}%
\end{equation}
where we have relaxed the quantities $\sigma_{\mathrm{a}}$ and $G$. Fig.
\ref{FIg_PDF} and even more \ref{Fig. PDF-LOG} show that, at least when the
amount of adhesion energy is small (as in our case), Eq.
(\ref{final double gaussian}) is a good approximation of the numerically
calculated stress probability function $P\left(  \sigma,\zeta_{\max}\right)
$. Indeed Fig. \ref{Fig. PDF-LOG} shows that, at high values of $\sigma$, the
tail of the stress probability density function is Gaussian. Notice that, the
tail of numerically calculated stress probability distribution at negative
loads is an effect of the adhesion interaction which has been introduced only
through the surface energy, namely by means of an interaction force with an
infinitesimally short range. This even allows that infinite negative values of
$\sigma$ can occur at the interface. We also observe that the value of
$\sigma_{\mathrm{a}}$ calculated by fitting the interfacial stress probability
density function differs from that calculated in the spirit of Persson's
theory through Eq. (\ref{apparent traction stress}). However, it is possible
to show that a better estimation of the tensile stress $\sigma_{\mathrm{a}}$,
can be obtained by assuming in Eq. (\ref{apparent traction stress}) that the
width of the detached region is a factor 1/2 smaller than that originally
assumed in Persson's theory \cite{PerssonAdhesion}. In such a case the
estimated tensile stress $\sigma_{\mathrm{a}}$ would increase of roughly a
factor 1.4 thus making $\sigma_{\mathrm{a}}$ closer to the value we have found
in Fig. \ref{FIg_PDF}.

\section{Conclusions}

The authors have carried out detailed numerical calculations to determine the
contact area, the stress distribution, the penetration and elastic deformation
of an infinitely thick layer in adhesive contact with a rough strongly
anisotropic rigid surface. The numerical predictions have been compared in
detail with those of an extended version of Persson's theory able to deal with
adhesive contact between anisotropic rough surfaces. It is shown that, for any
given load, the value of true contact area predicted by Persson's theory
significantly differs from the numerically calculated ones, the first being
smaller by a factor $\approx1/2$. This may also depend on the fact the
Persson's theory is of the mean-field type and, therefore, should work well in
higher dimensions than in 1D. However, the predicted value of separation
matches almost perfectly the numerical data. We have also compared the power
spectral density of the deformed elastic surface and of the stress
distribution at the interface as obtained by numerical calculations and
Persson's theory. We observe that both theory and numerical calculations
predict the PSDs to follow a power law with almost the same exponents. This
extends to the case of adhesive contact, what has been found previously for
adhesionless contact by other authors. However, we also observe, in agreement
with Persson's theory of adhesive contact, that at high magnification the
exponent of the power law changes in such a way to suggest that the elastic
solid lies in full contact condition with the fine microstructure of the rough
surfaces. We conclude that, Persson's theory is able to capture the main
physics behind contact mechanics of rough surfaces independently of whether
adhesion is present or not. However, we also observe that from an engineering
point of view a better estimation of the contact area would be very useful in
practical applications as in case tires, mixed lubricated interfaces, and
seals, where the kinetic friction or the amount of leakage should be
accurately predicted for design purposes. Therefore an improvement of the
theory, which allows to better estimate the real contact area, would be
strongly appreciated by engineering community.

\begin{acknowledgments}
This work, as part of the European Science Foundation EUROCORES Programme
FANAS was supported from the EC Sixth Framework Programme, under contract N. ERAS-CT-2003-980409
\end{acknowledgments}

\end{document}